# A second mapping method in generalized discrete singular convolution algorithm: regularizing singularities for one electron system

Kaige Hu[+] · Ruiqin Zhang*

*Department of Physics and Material Science, City University of Hong Kong, Hong Kong SAR, China*

[+] Present address: International Center for Quantum Materials, Peking University, Beijing, China
* Corresponding author: E-mail: aprqz@cityu.edu.hk

**Abstract** A second mapping method is introduced in the generalized discrete singular convolution algorithm. The mapping approaches are adopted to regularize singularities for one electron system. The applications of the two mapping methods are generalized from the radial hydrogen problem to the one-dimensional hydrogen problem. Three mapping functions are chosen: the square-root mapping function, the cube-root mapping function, and the logarithm mapping function. However, the present mapping approaches fail in both the two-dimensional and three-dimensional hydrogen problems, because the wavefunctions of $s$-states at the nuclei are not correct.

## 1 Introduction

Many numerical algorithms have been introduced to sovle the Schrödinger equation in quantum mechanics and the Kohn-Shan equation in density functional theory (DFT). For example, the discrete variable representation (DVR) method, the Lagrange-mesh method, the finite different (FD) methods, the finite element (FE) methods, the wavelet method, etc. Since the singularity of the Coulomb potentials, whether a given algorithm is feasible or not depends on its viablity in the presence of such singularities. Why singularities usually introduce numerical instabilities? It can be intuitively understood by that the discrete sampling in numerical approaches fails to be a good representation of the original continuous term because the changes near the singularity are very sharp. Several methods can deal with the Coulomb singularity by their original algorithms, such as the path-integral quantum Monte Carlo method [1], the asymptotic iteration method [2,3], the hyperspherical coordinates [4], the Ritz method [5,6], and the DVR using the Lagrange-Lobatto basis [7]. However, it is natural that a method needs regularization in the presence of singularities. For example, the DVR method using the lagrange basis provides poor results in the presence of the Coulomb or centrifugal potentials [8], and it turns out that the singular difficulties can be removed by multiplying a positive scaled function to the left of the equation and producing a modified Hamiltonian [8-13]. In the Fourier grid method [14-16], a new form of the Hamiltonian is obtained since the mapping operates on the singular potential, and the grid points are still evenly distributed after the mapping. Such grid-transformation approaches have been generalized to numerical integration of exchange-correlation energies and potentials in DFT [17,18]. However, since many numerical methods can only give the accurate ground state or few excited states, or are difficulty to construct the Hamiltonian matrix when the number of basis or grid points increase, the development of incoming numerical algorithms has never been stopped.

In recent years, a discrete singular convolution (DSC) algorithm has been proposed [19] and applied in various engineering systems without singularities and obtained competitive results [20-25]. Therefore we have been encouraged to apply the DSC in one-electron physical systems [26,27]. The DSC has been used to solve the radial Schrödinger equation of a hydrogen atom [26]. However, since the results in [26] are affected distinctly by the Coulomb singularity, we introduced a mapping method to obtain a nonuniform discretization to regularize the singularity [27]. The mapping procedure is used to produce a nonuniform discretization which is appropriate to the singular potential from a uniform discretization, but with the form of the potential and Hamiltonian unchanged. The new DSC kernels are constructed from the new basis functions, which are mapped from the original basis functions.

In this paper, a second mapping method was introduced in the generalized DSC (GDSC) algorithm. The two mapping approaches obtain non-uniform discretization from uniform discretization. Different from that in the first mapping method, the new kernels in the second mapping method are mapped from the original kernels directly. The elements of the Hamilton matrix are expected to be very simple when the inverse function of the mapping function is an explicit function. The programs of the GDSC are very short and it is convenient to consider different numbers of grid points and even a very large one. The regularization of singularities for one electron system are explored. The applications of the two mapping methods are generalized from the radial hydrogen problem to the one-dimensional hydrogen problem. Three mapping functions are chosen: the square-root mapping function, the cube-root mapping function, and the logarithm mapping function. The GDSC can produce excellent eigenvalues for both the ground state and many excited states.

## 2 Principle

### 2.1 Generalized discrete singular convolution algorithm

A singular convolution (SC) is defined as

$$f(x) = (T * g)(y) = \int T(x, y) g(y) dy, \qquad (1)$$

where $T(x, y)$ is a singular kernel. Usually the kernel only depends on the difference of the two variables $x$ and $y$ and is described as $T(x - y)$. However, here we extend the expression of a kernel to a generalized case because a nonuniform discretization is considered. There are several types of singular kernel, such as the Hilbert transform, Abel transform, Radon transform and delta transform [20]. In the application to physical systems, we are interested in the singular kernels of the *delta type*

$$T(x, x') = \delta^{(n)}(x, x') \equiv \frac{\partial^n \delta(x, x')}{\partial x^n}, n = 0,1,2,\ldots, \quad n = 0,1,2,\ldots \qquad (2)$$

The kernel plays key roles in numerically solving partial differential equations. For the kernel, *differential operators always* operate on the first variable, while the integrals are all about the second variable. When delta type kernels are taken, we can approximate a given function and also its derivatives

$$f^{(n)}(x) = \int \delta^{(n)}(x - x') f(x') dx'.$$

Apply the SC with singular kernels of the delta type to the Schrödinger equation $H\psi(x) = E\psi(x)$, we obtain

$$\int H(x, x') \psi(x') dx' = E\psi(x) \qquad (3)$$

where $H$ is the Hamiltonian operator, and $\psi(x)$ is the wavefunction corresponding to energy $E$. For a typical Hamiltonian $H = T + V = -\frac{1}{2} d^2/dx^2 + V(x)$, where $T$ and $V$ are the kinetic operator and potential operator, respectively, we obtain

$$H(x,x') = -\frac{1}{2}\delta^{(2)}(x-x') + V(x)\delta_{x,x'} \tag{4}$$

For numerical purpose, discretization for SC is needed, i.e., DSC must be considered. However, since the kernels are singular, they cannot be used directly in numerical computation. Thus sequences of approximations $\delta_\alpha(x-x')$ must be constructed

$$\delta(x,x') = \lim_{\alpha \to \alpha_0} \delta_\alpha(x,x'), \tag{5}$$

where $\alpha_0$ is a generalized limitation. For example, the Shannon kernel is described as

$$\delta_\alpha(x,x') = \frac{\sin[\alpha(x-x')]}{\pi(x-x')}, \tag{6}$$

with the generalized limitation being $\alpha_0 = \infty$. It is an instance for the uniform discretization. The limitation of Shannon kernel satisfies the definition of delta function over the interval $(-\infty, \infty)$.

Now the DSC can be considered. The original DSC adapted by Wei [31,36,32,35] is with uniform discretization. In this paper a generalized DSC (GDSC) is adapted, i.e., nonuniform discretizations are considered:

$$f_\alpha(x) = \sum_i \lambda_i T_\alpha(x, y_i) g(y_i) \tag{7}$$

where $f_\alpha(x)$ is an approximation to $f(x)$, $y_i$ is an appropriate set of discrete points on which DSC is well defined, and $\lambda_i$ are the corresponding weights. $\lambda_i$ are the Christoffel numbers associated with a Gauss quadrature formula

$$\int g(y) dy \approx \sum_j \lambda_j g(y_j) \tag{8}$$

Specially the GDSC for the wavefunction of the Schrödinger equation reads

$$\psi(x) = \sum_j \lambda_j \delta_\alpha(x, x_j) \psi(x_j) = \sum_j \widetilde{K}_\alpha(x, x_j) \psi(x_j) \tag{9}$$

where $\widetilde{K}_\alpha(x,x_j) \equiv \lambda_j \delta_\alpha(x,x_j)$ is the discrete version of a given singular kernel. The matrix-eigenvalue problem for the Schrödinger equation becomes

$$\sum_j \widetilde{H}(x_i,x_j)\psi(x_j) = E\psi(x_i) \tag{10}$$

where

$$\widetilde{H}(x_i,x_j) = -\frac{1}{2}\widetilde{K}_\alpha^{(2)}(x_i,x_j) + V(x_i)\delta_{ij} \tag{11}$$

From (10) and (11), the eigenenergies can be obtained numerically by diagonallizing the Hamiltonian matrix directly. Mathematically, (10) has no problem to provide right eigenvalues (eigenenergies). However, the normalized eigenvectors do not correspond to normalized wavefunctions physically. A normalized eigenvector satisfies the relation $\sum_j |\psi(x_j)|^2 = 1$, but a normalized wavefunction should satisfy the relation

$$\int |\psi(x)|^2 dx \approx \sum_i \lambda_i |\psi(x_i)|^2 = \sum_i |\lambda_i^{1/2}\psi(x_i)|^2 = 1 \tag{12}$$

(12) hints that the *discrete wavefunction* is not $\psi(x_i)$ but $\lambda_i^{1/2}\psi(x_i)$! Thus the GDSC for the Schrödinger equation can be modified to

$$c_i = \sum_j K_\alpha(x_i,x_j)c_j, \tag{13}$$

with the new discrete version of the kernel given by

$$K_\alpha(x_i,x_j) = (\lambda_i\lambda_j)^{1/2}\delta_\alpha(x_i,x_j). \tag{14}$$

(14) is the relationship between the continuous and discrete kernels. After the discretization, the matrix-eigenvalue problem of the Schrödinger equation becomes

$$\sum_j H(x_i,x_j)c_j = Ec_i, \tag{15}$$

where $H(x_i,x_j)$ is the matrix element for the Hamiltonian operator $H$

$$H(x_i,x_j) = -\frac{1}{2}K_\alpha^{(2)}(x_i,x_j) + V(x_i)\delta_{ij}. \tag{16}$$

Where from (14) we know that

$$K_\alpha^{(2)}(x_i,x_j) = (\lambda_i\lambda_j)^{1/2}\delta_\alpha^{(2)}(x,x_j)|_{x=x_i} \tag{17}$$

The continuous wavefunction at any position can be calculated as

$$\psi(x) = \sum_j \lambda_j^{1/2} \delta_\alpha(x, x_j) c_j. \tag{18}$$

Since eigenenergies given by (10) and (15) are almost the same to each other and (15) is much more physical than (10), we will always adopt (15) in our studies.

In the DSC approach, a grid representation for the coordinate is chosen, so that the potential part $V(x)$ of the Hamiltonian is diagonal. Actually (13) shows that

$$K_\alpha(x_i, x_j) = \delta_{ij}, \tag{19}$$

which means that the discrete delta-type kernel should be the Kronecker delta. It is reasonable because the Kronecker delta is an analogue of the Dirac delta function with discrete indices. Equivalently, the continuous delta-type kernel $\delta_\alpha(x, x_j)$ should satisfy

$$\delta_\alpha(x_i, x_j) = \lambda_j^{-1} \delta_{ij}. \tag{20}$$

This relation shows that the parameter $\alpha$ is related to the discrete grid points. For a given set of grid points which represents the continuous variable, the parameter $\alpha$ which represents the approximation of the delta function is a specific value. Concrete examples will be studied in the coming sections.

For the calculation of weights $\lambda_i$ to grid points $x_i$, usually a set of orthonormal basis functions $\varphi_k(x)$ are needed

$$\lambda_i = \left[ \sum_k |\varphi_k(x_i)|^2 \right]^{-1}. \tag{21}$$

The grid points $x_i$ are zero points of a reference basis function $\psi_N(x)$, where $N$ is the number of grid points. However, the case is much simpler for uniform discretization because the grid points are evenly distributed.

Here is an interpretation to understand the excellent accuracy of the DSC. The DSC can provide an approximate wavefunction at any $x$ by (18) and not only at the discrete grid points $x_i$. Such advantages result from the *global* feature of the kernel $\delta_\alpha(x, x')$, although the strict delta function $\delta(x, x')$ is *local* strictly. Notice that the *local* $\delta(x, x')$ is a generalized function, whose exact meaning must be understood under integrals as the limitation of some sequence of *global* functions $\delta_\alpha(x, x')$. The DSC method links the *local* and *discrete* grid points with the *global* and *continuous* wavefunction naturally, and therefore it can provide accuracy eigenenergies.

Although the one-dimensional Schrödinger equations is used as an example, multi-dimensional cases and any other eigen equations can be treated similarly. For example, a trivial two-dimensional Hamiltonian is described as $H = -\frac{1}{2}\left(\partial^2/\partial x^2 + \partial^2/\partial y^2\right) + V(x,y)$, and the Hamiltonian-DSC matrix elements are given by

$$H_{ij} = H[(x_i, x_j), (y_i, y_j)]$$
$$= -\frac{1}{2}\left[K^{(2)}(x_i, x_j)K(y_i, y_j) + K(x_i, x_j)K^{(2)}(y_i, y_j)\right] + V(x_i, y_i)\delta_{ij} \quad (22)$$

where $i$ are the index of grid points, $(x_i, y_i)$ are the corresponding coordinates. The following relation has already been used

$$\delta(x-x', y-y') = \delta(x-x')\delta(y-y') \quad (23)$$

(23) indicates that the construction of multi-dimensional delta type kernel is variable separable, which is a great advantage of the DSC algorithm. Moreover, the two-dimensional Hamiltonian matrix is an *sparse* matrix since $H_{ij} = 0$ for any $i_x \neq j_x, i_y \neq j_y$,, while $i_x(i_y)$ are the index of the grid points in $x(y)$-direction. Similarly any multi-dimensional Hamiltonian matrices are *sparse*, which is a huge advantage because the calculation of eigenenergies for sparse matrices can be much simpler than full matrices.

## 2.2 The uniform discretization as a special case

There are several kernels suitable for uniform discretization, such as the Shannon kernel, the Dirichlet kernel, the modified Dirichlet kernel and the de la Vallée Poussion kernel. We take the Shannon kernel as an example of uniform discretization. The weights $\lambda_i$ are all equal to the grid interval $\Delta$, i.e.,

$$\lambda_i = \Delta. \quad (24)$$

Thus the Shannon kernel is discretized into

$$K_\alpha(x_i, x_j) = \frac{\sin[\alpha(x_i - x_j)]}{(\pi/\Delta)(x_i - x_j)}. \quad (25)$$

The parameter $\alpha$ should be chosen so that (19) is satisfied. Noticing $x_i - x_j = (i-j)\Delta$, it is obvious that

$$\alpha = \frac{\pi}{\Delta}, \tag{26}$$

which is the *Nyquist frequency* in the sampling theorem. This relationship between $\alpha$ and $\Delta$ is adapted in most of the previous applications of DSC. The second derivative of $K_\alpha(x_i, x_j)$ can be expressed by compact expressions:

$$K_\alpha^{(2)}(x_i, x_j) = \begin{cases} -\frac{\alpha^2}{3} & (i = j) \\ (-1)^{i-j+1} \frac{2}{(x_i - x_j)^2} & (i \neq j) \end{cases} \tag{27}$$

In practical calculations, the number of grid points $N$ must be finite, i.e., a truncation must be taken. Such a truncation slightly destroys the definition of the continuous kernel since the integral of $\delta_\alpha(x, x')$ is 1 only for the interval $(-\infty, \infty)$. However, the discrete kernel $K_\alpha(x_i, x_j)$ is not destroyed and remains as the Kronecker delta $\delta_{ij}$, which is reasonable. The truncation is nothing but abandoning the intervals which are not important to the required eigenstates, and should have no effect on the interval which is kept. Physically, the potentials on the abandoned intervals are set to be infinity and therefore the values of the wavefunction are forced to be zero [26,27].

## 2.3 Nonuniform discretization

Generally for nonuniform discretization, the problem is how to construct suitable kernels for a given set of grid points $x_i$ and weights $\lambda_i$. Since $x_i$ and $\lambda_i$ are defined by a set of orthonormal basis functions $\varphi_k(x)$, we may expect that the continuous kernel $\delta(x, x')$ and the discrete kernel $K(x_i, x_j)$ can be constructed from the basis $\varphi_k(x)$ too. Correlative techniques have already been developed in the DVR and the Lagrange-mesh method. There are two different ways to build up delta-type kernels, which are similar to building up the Lagrange functions in the Lagrange-mesh method. The first approach expands the kernel on a set of orthonormal basis functions $\varphi_k(x)$. The continuous kernel can be constructed as [8]

$$\delta_\alpha(x, x_j) = \sum_{k=1}^{N} \varphi_k^*(x_j) \varphi_k(x), \tag{28}$$

which satisfies (20) at the grid points $x = x_i$. The discrete kernel is given by

$$K_\alpha(x_i, x_j) = (\lambda_i \lambda_j)^{1/2} \sum_{k=1}^{N} \varphi_k^*(x_j) \varphi_k(x_i), \tag{29}$$

which satisfies (19). The second is the compact expressions of the delta kernels, as introduced by Schwartz [28]. The continuous kernel can be constructed as [8]

$$\delta_\alpha(x, x_j) = \lambda_j^{-1} \frac{\varphi_N(x)}{\varphi'_N(x_j)} \frac{1}{x - x_j}, \tag{30}$$

which satisfies (20) at the grid points $x = x_i$. It is obvious that the continuous kernel (30) satisfies the definition of $\delta$-function when the Gauss approximation is used

$$\lim_{N \to \infty} \left( \lim_{x \to x_j} \delta(x, x_j) \right) = \infty, \tag{31}$$

$$\int \delta(x, x_j) dx = 1. \tag{32}$$

The discrete kernel is given by

$$K_\alpha(x_i, x_j) = (\frac{\lambda_i}{\lambda_j})^{1/2} \frac{\varphi_N(x_i)}{\varphi'_N(x_j)} \frac{1}{x_i - x_j}, \tag{33}$$

which satisfies (19), i.e., it is indeed the Kronecker delta. Since the compact expressions of delta kernels can provide compact expressions for the Hamiltonian matrix elements, we will adapt the second approach in the calculations of this work.

## 2.4 Mapping from uniform to nonuniform discretization

Since the DSC with uniform discretization performs perfectly for problems without singularities, one only need consider non-uniform discretization for singular problems. In this section, we introduce a mapping method to obtain non-uniform grid points from uniform grid points. In order to regularize singularities, the mapping function must be chosen carefully so that the grid points are denser near the singular points and sparser in other regions for a given problem. When the inverse function of the mapping function has an analytical expression, the coordinates of non-uniform grid points also have analytical expressions. Therefore the matrix elements of the Hamiltonian are more concise and the numerical programs are simpler. Furthermore, one just needs to *simply* change the parameters to consider a different number of grid $N$ (even a very large one).

Consider a Schrödinger equation on the interval $x \in [0, L]$, with the boundary conditions

$$\psi(x=0) = \psi(x=L) = 0. \tag{34}$$

It is easy to construct a set of orthogonal sine basis

$$\varphi_k(x) = \sqrt{\frac{2}{L}} \sin\left[\frac{(k+1)\pi}{L} x\right], \quad k=1,2,...,N. \tag{35}$$

Especially for $k = N$

$$\varphi_N(x) = \sqrt{\frac{2}{L}} \sin\left[\frac{\pi}{\Delta} x\right], \tag{36}$$

where

$$\Delta = \frac{L}{N+1}. \tag{37}$$

Grid points are given by

$$x_i = i\Delta. \tag{38}$$

Weights $\lambda_i$ and parameter $\alpha$ are given by (21) and (26), respectively. The continuous and kernel is given by

$$\delta_\alpha(x, x_j) = \frac{\sin[\alpha(x - x_j)]}{\pi(x - x_j)}, \tag{39}$$

and the discrete kernel is given by

$$K_\alpha(x_i, x_j) = \frac{\sin[\alpha(x_i - x_j)]}{\alpha(x_i - x_j)}, \tag{40}$$

which are just the Shannon kernel. The second derivative of $K_\alpha(x_i, x_j)$ is given by (27).

### 2.4.1. The first mapping method

Now introducing a mapping function $X = g(x)$, and we have $x = g^{-1}(X)$ and $g'(x) = dX/dx$. The aim of such a mapping procedure is to obtain a new orthogonal basis and also new delta-type kernels, which are suitable for non-uniform discretization in the new variable $x$. Note that we do not map the Schrödinger equation from $X$ to $x$ at the same time but consider it directly on the variable $x$, which avoids nontrivial expressions caused by the mapping. Therefore the new matrix-eigenvalue problem and the matrix elements of $H$ can be obtained simply by replacing $X$ by $x$ in (15) and (16), respectively.

The sine basis (35) is mapped to a new orthogonal sine basis

$$\phi_k(x) = \sqrt{g'(x)} \psi_k(X(x)) = \sqrt{\frac{2}{L}} \sin\left[\frac{(k+1)\pi}{L} g(x)\right], \quad k = 1, 2, \ldots, N. \quad (41)$$

The region for the new variable $x$ is $[0, g^{-1}(L)]$. Note that there is no weight function for the new basis because the term $g'(x)$ is included in the expression of $\phi_k(x)$. Specially for $k = N$

$$\phi_N(x) = \sqrt{g'(x)} \sin[\alpha g(x)] \quad (42)$$

The grid points are zero points of $\phi_N(x)$ which are given by

$$x_i = g^{-1}(X_i) = g^{-1}(i\Delta) \quad (43)$$

It is obvious that $\phi_N(x_i) = 0$. The weights are given by

$$\lambda_i = \frac{\Delta}{g'(x_i)}, \quad (44)$$

which is calculated by (21). The continuous kernel is

$$\delta_\alpha(x, x_j) = \sqrt{\frac{g'(x)}{g'(x_j)}} \frac{\sin[\alpha(g(x) - g(x_j))]}{\pi(x - x_j)}, \quad (45)$$

and the discrete kernel is

$$K_\alpha(x, x_j) = \left[\frac{1}{g'(x_j)} \sqrt{\frac{g'(x)}{g'(x_i)}} \frac{\sin[\alpha(g(x) - g(x_j))]}{\alpha(x - x_j)}\right]_{x = x_i} \quad (46)$$

Note that generally the parameter $\Delta$ is not the grid interval any more since the discretization is non-uniform. The expressions of $x_i$, $K_\alpha(x_i, x_j)$ and even $K_\alpha^{(2)}(x_i, x_j)$ are expected to be much simpler than the general expression when the inverse function of the mapping function $X = g(x)$ is an explicit function. In general, the Hamiltonian matrix is not Hermitian and therefore standard non-Hermitian or non-symmetric diagonalization procedures must be used.

### 2.4.2 The second mapping method

In the former mapping method, the new kernels are constructed from the new basis functions, which are mapped from the original sine basis functions. However, there is an alternative choice. The main difference of the second approach from the first one is that the new kernels are mapped from the original kernels directly. Suppose the original delta-type kernel for the uniform discretization is the Shannon kernel $\delta_\alpha(X - X')$, satisfying

$$\int \delta_\alpha(X-X_j)dX = 1 = \int \delta_\alpha(g(x)-g(x_j))g'(x)dx. \tag{47}$$

$$\delta_\alpha(x,x_j) = g'(x)\delta_\alpha(g(x)-g(x_j)). \tag{48}$$

and the discrete kernel is given by (14). Therefore for a trivial Hamiltonian, the matrix element is given by (16).

## 3 Applications

### 3.1 The radial Schrödinger equation for hydrogen atom

In this section, we consider both the two types of mapping methods and consider three different mapping functions: the square-root mapping function

$$X = x^{1/2}, \tag{49}$$

the cube-root mapping function

$$X = x^{1/3}, \tag{50}$$

and the logarithm mapping function

$$X = \ln(1+x). \tag{51}$$

Table 1 lists the common parameters of the three different mapping functions and Table 2 gives the second order discrete kernel $K_\alpha^{(2)}(x_i, x_j)$. Figure 1 is the Curves of the three different mapping functions. The results are given in Tables 3 and 4. It turns out that both the two mapping methods work dramatically. Among them the cube-root mapping works the best, then the square-root mapping and finally the logarithm mapping. It is reasonable since the non-uniform grid points generated by cube-root mapping are the densest near the singular point $x=0$, and that generated by logarithm mapping are the sparsest among the three, which is obvious in Figure 1. An interesting phenomenon is that the results of logarithm mapping under the two different mapping methods are identical to each other, although their Hamiltonian matrix elements are different from each other (see Table 2).

**Table 1.** Parameters of the three different mapping functions: the square-root mapping function $g(x) = x^{1/2}$, the cube-root mapping function $g(x) = x^{1/3}$ and the logarithm mapping function $g(x) = \ln(1+x)$.

| | Square-root mapping | Cube-root mapping | Logarithm mapping |
|---|---|---|---|
| $g(x)$ | $x^{1/2}$ | $x^{1/3}$ | $\ln(1+x)$ |
| $g'(x)$ | $\dfrac{1}{2x^{1/2}}$ | $\dfrac{1}{3x^{2/3}}$ | $\dfrac{1}{1+x}$ |
| $g''(x)$ | $-\dfrac{1}{4x^{3/2}}$ | $-\dfrac{2}{9x^{5/3}}$ | $-\dfrac{1}{(1+x)^2}$ |
| $\Delta$ | $\dfrac{\sqrt{R}}{N+1}$ | $\dfrac{\sqrt[3]{R}}{N+1}$ | $\dfrac{\ln(1+x)}{N+1}$ |
| $X_i$ | $i\Delta$ | $i\Delta$ | $i\Delta$ |
| $x_i = g^{-1}(X_i)$ | $i^2\Delta^2$ | $i^3\Delta^3$ | $e^{X_i} - 1$ |
| $\lambda_i = \Delta / g'(x_i)$ | $2i\Delta^2$ | $3i^2\Delta^3$ | $\Delta(1+x_i)$ |

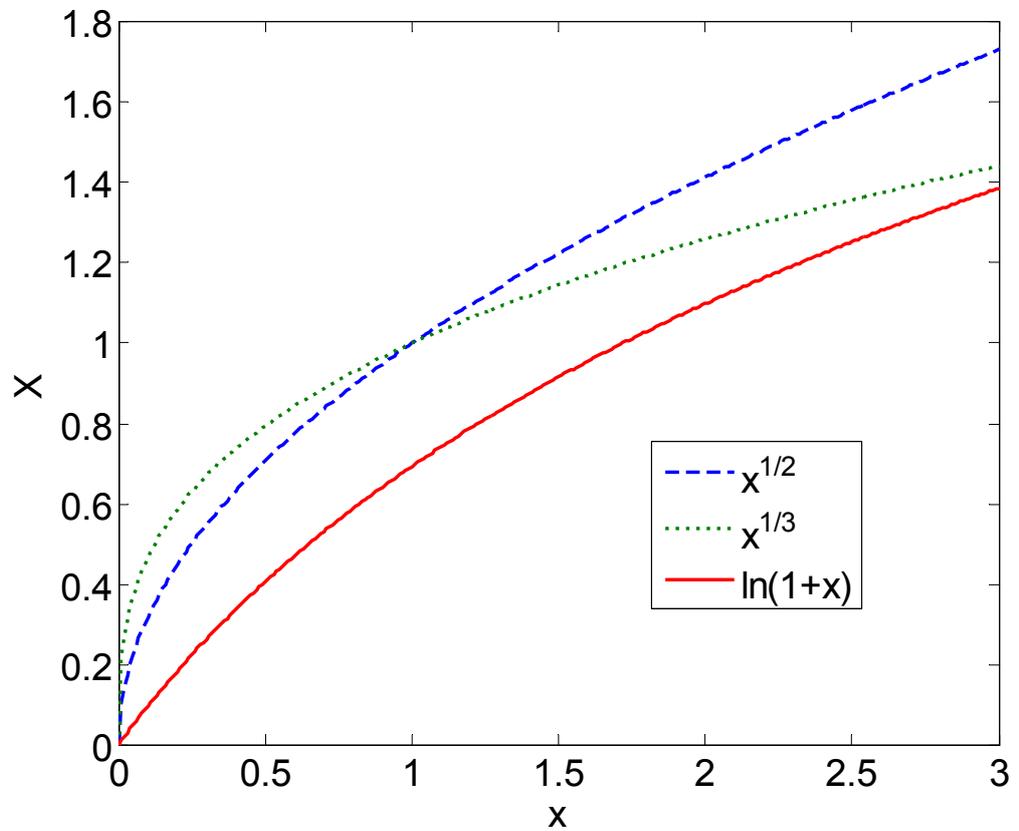

**Figure 1.** Curves of the three different mapping functions: the square-root mapping function, the cube-root mapping function and the logarithm mapping function.

**Table 2.** $K_\alpha^{(2)}(x_i, x_j)$ for the two different mapping methods and the three different mapping functions, respectively. The mapping functions are (1) the square-root mapping function $g(x) = x^{1/2}$, (2) the cube-root mapping function $g(x) = x^{1/3}$ and (3) the logarithm mapping function $g(x) = \ln(1+x)$. Parameters used are $\overline{E} = E_5$, $R = 50$, and $N = 500$. The powers of ten are indicated in square brackets.

| $K_\alpha^{(2)}(x_i, x_j)$ | | Square-root mapping ($x^{1/2}$) | Cube-root mapping ($x^{1/3}$) | Logarithm mapping ($\ln(1+x)$) |
|---|---|---|---|---|
| The 1st mapping method | $i \neq j$ | $\dfrac{(-1)^{i-j+1} j(5i^2 - j^2)}{i^3(i^2 - j^2)^2 \Delta^4}$ | $\dfrac{(-1)^{i-j+1} 2j^2(5i^3 - 2j^3)}{3i^5(i^3 - j^3)^2 \Delta^6}$ | $(-1)^{i-j+1}$ $\times \dfrac{2(1 + 2x_i - x_j)(1 + x_j)}{(1 + x_i)^2 (x_i - x_j)^2}$ |
| | $i = j$ | $\dfrac{33 - 4i^2 \pi^2}{48 i^4 \Delta^4}$ | $\dfrac{28 - i^2 \pi^2}{27 i^6 \Delta^6}$ | $\dfrac{23 - 4\alpha^2}{12(1 + x_i)^2}$ |
| The 2nd mapping method | $i \neq j$ | $(-1)^{i-j+1} \dfrac{(ij)^{1/2}(5i - 3j)}{4(i - j)^2 (i\Delta)^4}$ | $(-1)^{i-j+1} \dfrac{2(4i - 3j)j}{9(i - j)^2 (i\Delta)^6}$ | $(-1)^{i-j+1} \sqrt{(1 + x_i)(1 + x_j)}$ $\times \dfrac{(2 + 3(i - j)\Delta)}{(1 + x_i)^3 (i - j)^2 \Delta^2}$ |
| | $i = j$ | $\dfrac{9 - (i\pi)^2}{12(i\Delta)^4}$ | $\dfrac{30 - (i\pi)^2}{27(i\Delta)^6}$ | $\dfrac{6 - \alpha^2}{3(1 + x_i)^2}$ |

**Table 3.** Absolute errors in the eigenenergies (a.u.) given by the 1st mapping DSC method with uniform discretization and the three different mapping functions: the square-root mapping function, the cube-root mapping function and the logarithm mapping function. Only the case for $l = 0$ is considered. Parameters used are $\overline{E} = E_5$, $R = 50$, and $N = 500$. The powers of ten are indicated in square brackets.

| $n$ | Exact $E_n$ | Original uniform case $X = x$ | Square-root mapping $X = \sqrt{x}$ | Cube-root mapping $X = \sqrt[3]{x}$ | Logarithm mapping $X = \ln(1+x)$ |
|---|---|---|---|---|---|
| 1 | $-0.50000$ | 2.2[−2] | 1.5[−4] | 1.0[−6] | 1.9[−3] |
| 2 | $-0.12500$ | 2.8[−3] | 1.9[−5] | 1.8[−6] | 2.3[−4] |
| 3 | $-0.05556$ | 8.3[−4] | 5.5[−6] | 4.5[−3] | 6.9[−5] |
| 4 | $-0.03125$ | 4.0[−4] | 3.8[−6] | 1.1[−2] | 7.6[−5] |
| 5 | $-0.02000$ | 2.4[−3] | 2.2[−3] | 2.2[−2] | 2.2[−3] |

**Table 4.** Absolute errors in the eigenenergies (a.u.) given by the 2nd mapping DSC method with uniform discretization and the three different mapping functions: the square-root mapping function, the cube-root mapping function and the logarithm mapping function. Only the case for $l = 0$ is considered. Parameters used are $\bar{E} = E_5$, $R = 50$, and $N = 500$. The powers of ten are indicated in square brackets.

| | | Original uniform case | Square-root mapping | Cube-root mapping | Logarithm mapping |
|---|---|---|---|---|---|
| $n$ | Exact $E_n$ | $X = x$ | $X = \sqrt{x}$ | $X = \sqrt[3]{x}$ | $X = \ln(1+x)$ |
| 1 | $-0.50000$ | 2.2[−2] | 9.0[−5] | 1.0[−6] | 1.9[−3] |
| 2 | $-0.12500$ | 2.8[−3] | 1.1[−5] | 2.1[−6] | 2.3[−4] |
| 3 | $-0.05556$ | 8.3[−4] | 1.1[−6] | 1.0[−3] | 6.9[−5] |
| 4 | $-0.03125$ | 4.0[−4] | 3.8[−5] | 2.3[−2] | 7.6[−5] |
| 5 | $-0.02000$ | 2.4[−3] | 2.5[−3] | 7.2[−3] | 2.2[−3] |

## 3.2 The One-dimensional Hydrogen Atom

For the 1D hydrogen atom problem only the logarithm mapping is feasible since infinities appear at the singular point $x = 0$ for square-root mapping and also cube-root mapping. Table 2 shows this feature clearly. When we generate the radial case to the 1D case, the indice $i$ can be 0 or negative. When it is 0, $K_\alpha^{(2)}(x_i, x_j)$ for square-root mapping or cube-root mapping diverges to infinity, which will also induce infinities in the Hamiltonian matrix and make the solution of the Hamiltonian matrix infeasible. Logarithm mapping avoids such infinities. Of course, we still have two approaches of mapping method. Table 5 shows the corresponding parameters. Note that the mapping functions are generalized from (51) to

$$X = g(x) = \text{sgn}(x)\ln(1+|x|), \tag{52}$$

which is an odd function. The length of the region considered is $L$, i.e., the region is $x \in (-L/2, L/2)$.

**Table 5.** Parameters of the logarithm mapping function.

| $X = g(x)$ | | $X = g(x) = \text{sgn}(x)\ln(1+|x|),\ x \in (-L/2, L/2)$ |
|---|---|---|
| $g'(x)$ | | $\dfrac{1}{1+|x|}$ |
| $g''(x)$ | | $-\dfrac{\text{sgn}(x)}{(1+|x|)^2}$ |
| $g^{(3)}(x)$ | | $\dfrac{2}{(1+|x|)^3}$ |
| $X_i$ | | $X = i\Delta,\ i = \ldots, -1, 0, 1, \ldots$ |
| $x_i = g^{-1}(X_i)$ | | $\text{sgn}(X_i)(e^{|X_i|} - 1)$ |
| $\lambda_i = \dfrac{\Delta}{g'(x_i)}$ | | $\Delta(1+|x_i|)$ |
| $K_\alpha^{(2)}(x_i, x_j)$ (1st mapping) | $i \neq j$ | $(-1)^{i-j+1} \dfrac{2(1+|x_i|) + \text{sgn}(x_i)(x_i - x_j)(1+|x_j|)}{(1+|x_i|)^2 (x_i - x_j)^2}$ |
| | $i = j$ | $\dfrac{23 - 4\alpha^2}{12(1+|x_i|)^2}$ |
| $K_\alpha^{(2)}(x_i, x_j)$ (2nd mapping) | $i \neq j$ | $(-1)^{i-j+1} \dfrac{(1+|x_j|)^{1/2}(2 + 3(X_i - X_j)\text{sgn}(x_i))}{(1+|x_i|)^{5/2}(X_i - X_j)^2}$ |
| | $i = j$ | $\dfrac{6 - \alpha^2}{3(1+|x_i|)^2}$ |

The results are shown in Table 6. Only the case for $l = 0$ is considered. Parameters used are $L = 60$ and $N = 60$. The performances of the two mapping methods are both dramatic: the errors are reduced by 1 or 2 orders.

**Table 6.** Absolute errors in the eigenenergies (a.u.) given by logarithm mapping by both the two mapping methods. Only the case for $l = 0$ is considered. Parameters used are $L = 60,$ and $N = 60$. The powers of ten are indicated in square brackets.

| $n$ | Exact $E_n$ | Uniform | 1$^{st}$ mapping | 2$^{nd}$ mapping |
|---|---|---|---|---|
| 1 | $-0.50000$ | 1.2[−1] | 3.2[−3] | 2.1[−3] |
|   |            | 1.8[−1] | 4.0[−2] | 4.0[−2] |
| 2 | $-0.12500$ | 1.6[−2] | 4.0[−4] | 2.6[−4] |
|   |            | 2.6[−2] | 5.2[−3] | 5.1[−3] |
| 3 | $-0.05556$ | 5.3[−3] | 9.4[−5] | 2.1[−4] |
|   |            | 9.0[−3] | 1.8[−3] | 1.8[−3] |
| 4 | $-0.03125$ | 1.2[−2] | 7.7[−3] | 8.5[−3] |
|   |            | 1.5[−2] | 8.6[−3] | 8.6[−3] |

In the calculation, the singular point (origin point $x = 0$) is included, i.e., it is located on a discrete grid point, where the potential is set as a large number $10^8$ to represent infinity. Such infinite Coulomb potential at the origin will be discussed in detail in the next section.

### 3.3 The Infinite Coulomb Potential at the Origin

For demonstration purposes, only the original uniform DSC needs to be considered.

The boundary conditions for the bond states are $\psi(x = \pm\infty) = 0$. Making a truncated approximation, we are considering a region $x \in (-L/2, L/2)$ and the boundary conditions become

$$\psi(x = \pm L/2) = 0. \tag{53}$$

We divide such region whose length is $L$ into $N$ pieces, i.e., the grid intervals are $\Delta = L/N$. There are $N+1$ grid points (including $x = \pm L/2$). However, since the values of the wavefunction at the two boundaries $x = \pm L/2$ are 0, we do not need to include them in our Hamiltonian matrix, i.e., we only need construct a $(N-1) \times (N-1)$ Hamiltonian matrix. The $N-1$ grid points are

$x_{-N/2+1}, ..., x_{-1}, x_0, x_1, ..., x_{N/2-1}$.

Figure 2 is a special case where $N=8$. A $7\times 7$ should be constructed. The seven grid points are $x_i = i\Delta$ where $i = -3,-2,-1,0,1,2,3$.

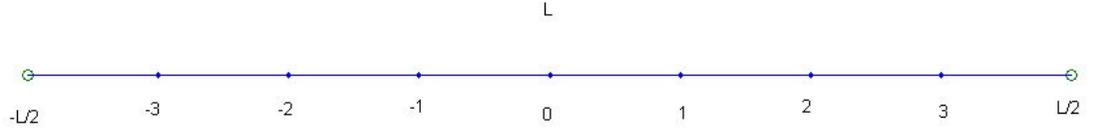

**Figure 2.** 1D grid on the interval $(-L/2, L/2)$, where $N=8$. The grid spacing is given by $\Delta = L/N$. The $i$th grid point is $x_i = i\Delta$ for $-N/2 < i < N/2$, i.e., $i = -3,-2,-1,0,1,2,3$.

Here is a question. Since $x=0$ is a singular point, should it be located on a grid point? If so, what should its potential value be? It is obvious that it will be located on a grid point when $N$ is even. Therefore, to avoid it, we should set $N$ as an odd number. It turns out that it is necessary to make $x=0$ located on a grid point, i.e., $N$ should be an even number. And it is better to set a positive infinite potential here. In real calculations, we use a large number, $10^8$, to represent such infinity. The detailed tests are provided in the following.

The limit of the Coulomb potential at $x=0$ is

$$\lim_{x\to 0} V(x) = \lim_{x\to 0} -\frac{1}{|x|} = -\infty, \tag{54}$$

which hints that setting a negative infinite potential here is reasonable. We use $-10^8$ to represent it. We tested three choices and the results are in Table 7. Another choice is using a positive infinite potential, $10^8$, and in this case all the eigenenergies are the same except abandoning a state with eigenenergy $-\infty(-10^8)$ (Figure 3). A third choice is avoiding $x=0$; however the even parity states become unacceptable since the infinite potential at $x=0$ is very important for such states (see Figure 4).

**Table 7.** Eigenenergies of three different choices of the potential as $x = 0$: (1) $V(0) = -10^8$; (2) $V(0) = 10^8$; (3) avoiding $x = 0$.

| Index $n$ | Exact $E_n$ | (1) $V(0) = -10^8$ (N=600, L=60) | (2) $V(0) = 10^8$ (N=600, L=60) | (3) avoid $x = 0$ (N=601, L=60) |
|---|---|---|---|---|
| 0 | $-\infty$ | $-10^8$ | - | -8.1790 |
| 1 | -0.5000 | -0.4970 | -0.4970 | -0.5016 |
|   | -0.5000 | -0.4662 | -0.4662 | -0.2988 |
| 2 | -0.1250 | -0.1246 | -0.1246 | -0.1252 |
|   | -0.1250 | -0.1207 | -0.1207 | -0.0949 |
| 3 | -0.0556 | -0.0553 | -0.0553 | -0.0555 |
|   | -0.0556 | -0.0541 | -0.0541 | -0.0454 |
| 4 | -0.0313 | -0.0245 | -0.0245 | -0.0247 |
|   | -0.0313 | -0.0235 | -0.0235 | -0.0148 |
| 5 | -0.0200 | 0.0135 | 0.0135 | 0.0133 |
|   | -0.0200 | 0.0150 | 0.0150 | 0.0273 |

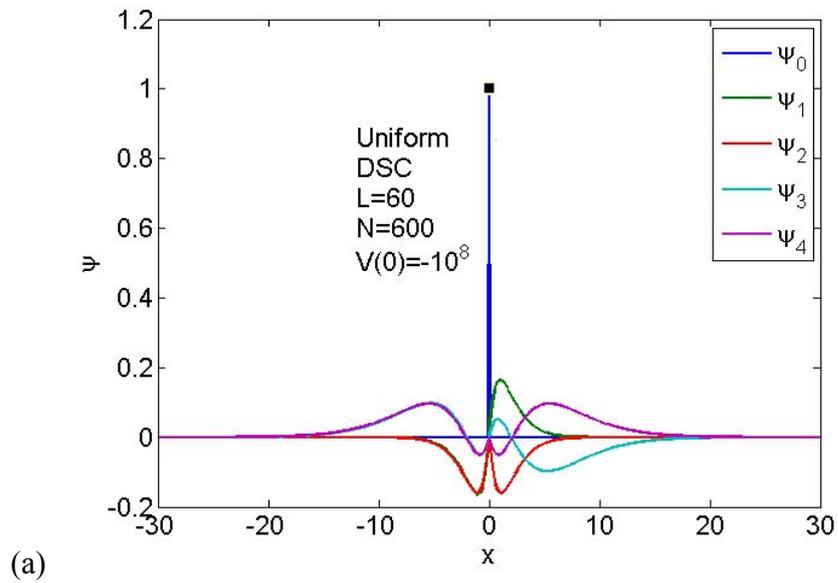

(a)

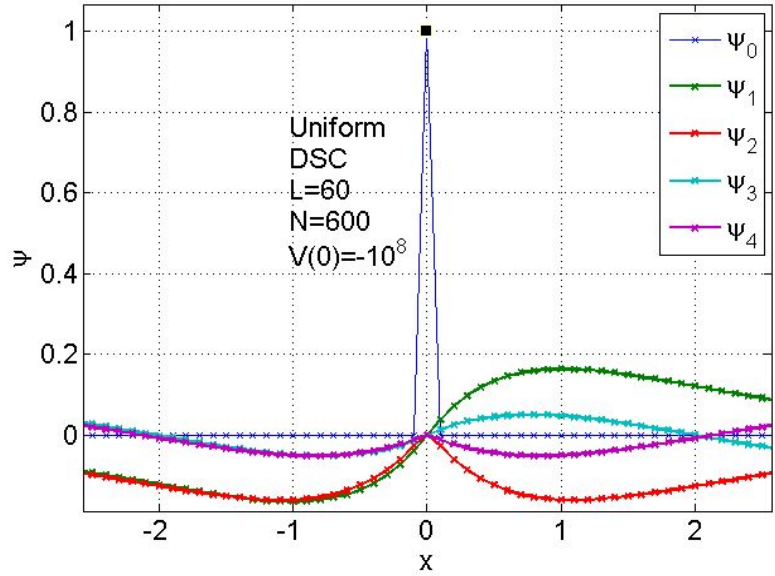

(b)

**Figure 3.** Wavefunctions with a negative infinite potential at the singular point $x=0$. Except the wavefunction with eigenenergy $-\infty$, all the wavefunctions, both even or odd parity wavefunctions, satisfy the boundary condition at $x=0$, i.e., $\psi(x=0)=0$. The wavefunction with eigenenergy $-\infty$ is located at $x=0$ and is 0 at any position $x \neq 0$. (a) A broad range; (b) The region near $x=0$.

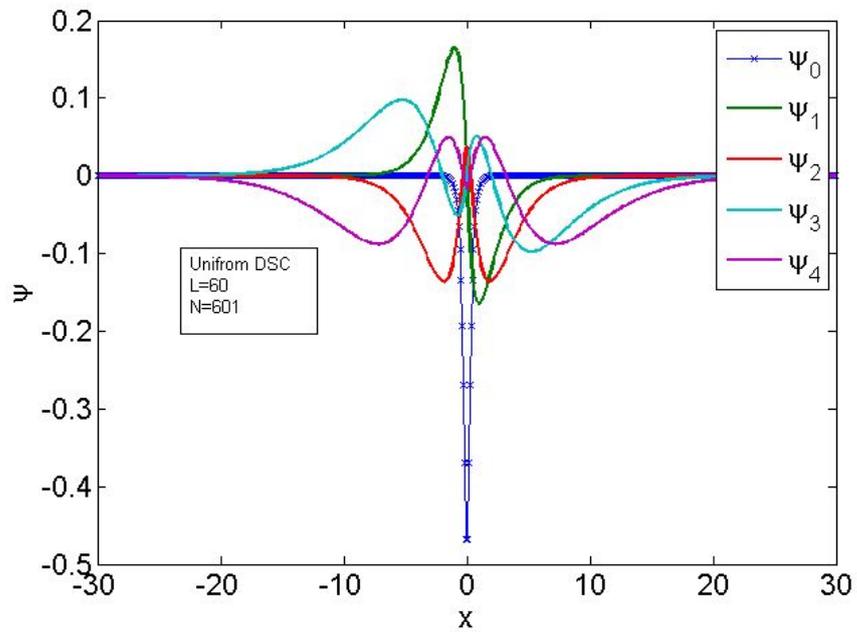

(a)

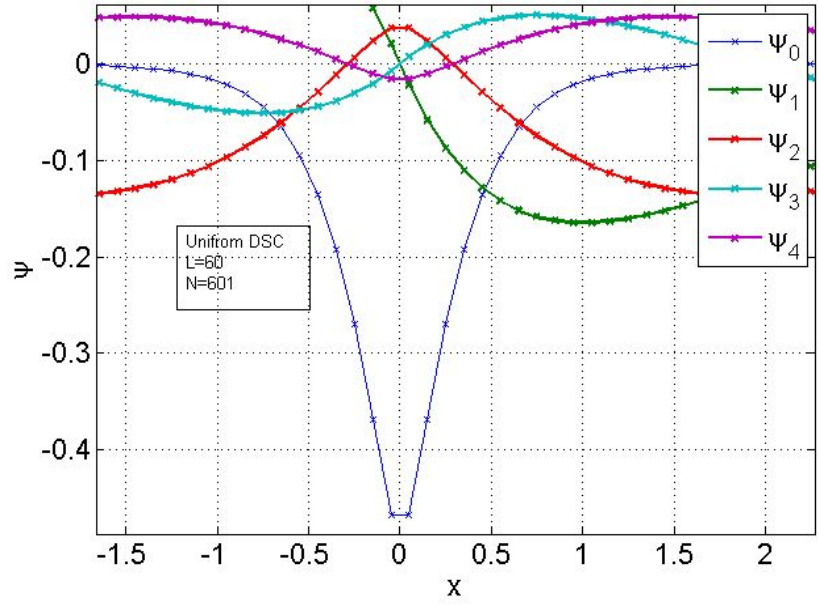

(b)

**Figure 4.** Wavefunctions without the singular point $x=0$. All the odd parity wavefunctions satisfy the boundary condition at $x=0$, i.e., $\psi(x=0)=0$. However, the even parity wavefunctions do not satisfy such condition. (a) A broad range; (b) The region near $x=0$.

How about changing the value of the "infinite" potential, i.e., using a value different from $10^8$? Figure 5 plots the five lowest eigenenergies versus $V(0)$, and the value is chosen as

$$V(0) = \text{sgn}(n)10^n, n = -8,-7,...,7,8. \tag{55}$$

It turns out that the absolute value must be large enough, or half of the eigenenergies will be unacceptable. These unacceptable states are with even parities, which can be checked by the corresponding wavefunctions. The difference between a positive infinity and negative infinity is an eigenstate with negative infinite eigenenergy. It is safe and reasonable to choose the positive infinity to abandon such a negative infinite eigenenergy.

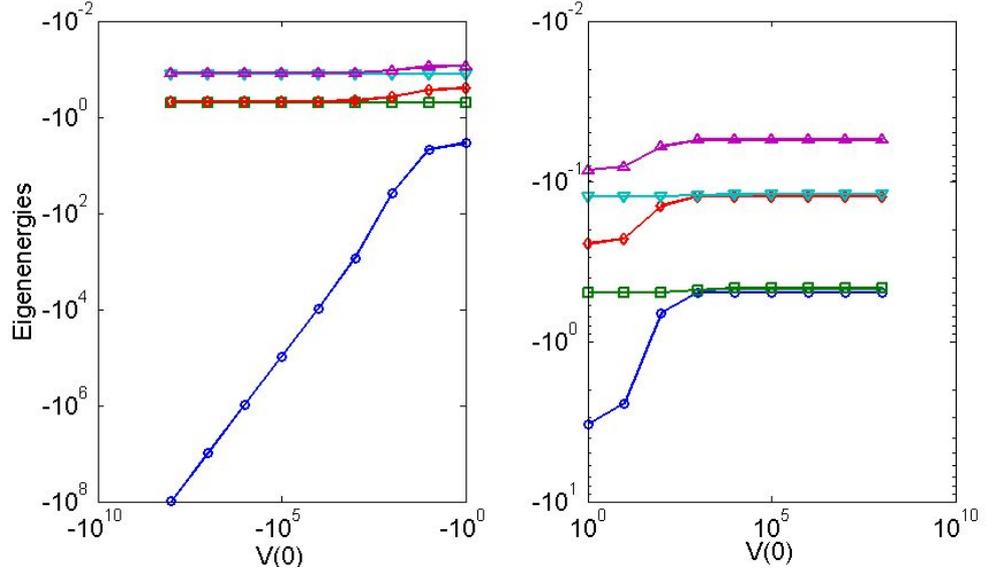

**Figure 5.** Plots of the five lowest eigenenergies versus $V(0)$ (the potential at the nuclear, i.e., $x = 0$).

*3.3 The two- and three-dimensional hydrogen atom*

Since delta-type kernels are variable separable, the 2D and 3D hydrogen atom problems are simple generations of the 1D case. Similarly, we set the singular point $\vec{r} = (0,0,0)$ at a grid point and set the Coulomb potential a positive infinite value.

Unfortunately, the two mapping methods both fail in both the 2D and 3D problems. As an example, in the following the results of the first mapping method for the 2D hydrogen atom are given in Table 8. The performance of the second mapping method for 2D hydrogen atoms and the two mapping methods for 3D hydrogen atoms are similar.

**Table 8.** Eigenenergies given by the first mapping method of DSC and the corresponding errors and relative errors. Parameters are chosen as $N_x = N_y = 40$, $L_x = L_y = 40$.

| Index $n$ | Exact $E_n$ | DSC $E_n$ | Error | Relative Error |
|---|---|---|---|---|
| 1 | -2.0000 | -0.8541 | 1.1459 | 0.5729 |
| 2 | -0.2222 | -0.3477 | 0.1255 | 0.5647 |
|   | -0.2222 | -0.3477 | 0.1255 | 0.5647 |
|   | -0.2222 | -0.2181 | 0.0041 | 0.0186 |
| 3 | -0.0800 | -0.1833 | 0.1033 | 1.2916 |
|   | -0.0800 | -0.1486 | 0.0686 | 0.8574 |
|   | -0.0800 | -0.1486 | 0.0686 | 0.8574 |
|   | -0.0800 | -0.1153 | 0.0353 | 0.4414 |
|   | -0.0800 | -0.1072 | 0.0272 | 0.3399 |
| 4 | -0.0408 | -0.0880 | 0.0472 | 1.1568 |

The most important reason that the mapping methods fail should be that the positive infinite potential at the singular point $\vec{r} = (0,0,0)$ forces the wavefunction there to be 0, which is obvious wrong for *s*-states.

## 4 Conclusions

A second mapping method was introduced in the GDSC algorithm. The two mapping approaches obtain non-uniform discretization from uniform discretization. In the first mapping method, the new kernels are constructed from the new basis functions, which are mapped from the original sine basis functions. However, in the second mapping method the new kernels are mapped from the original kernels directly. The elements of the Hamilton matrix are expected to be very simple when the inverse function of the mapping function is an explicit function. The programs of the GDSC are very short and it is convenient to consider different numbers of grid points and even a very large one. The regularization of singularities for one electron system was explored. The applications of the two mapping methods are generalized from the radial hydrogen problem to the one-dimensional hydrogen problem. Three mapping functions are chosen: the square-root mapping function, the cube-root mapping function, and the logarithm mapping function. The GDSC can produce excellent eigenvalues for both the ground state and many excited states. Since the GDSC method performs excellently in solving eigen equations with and without singularities, it is interesting to apply the GDSC method to more complicated systems, for example, $H_2^+$, *He* and other atoms and molecules, with or without electrical or magnetic fields.

However, up to now the GDSC has only been successful for the radial Schrödinger equation and the 1D hydrogen atom problem. It fails for the 2D and 3D hydrogen atoms since the infinite potential at the nuclear (origin) forces the values of the wavefunctions to be 0, which are not the cases for *s*-states. This situation indicates that new approaches are required to regularize singularities if we want to generate the application of the GDSC broadly.Since the singular Coulomb potentials influence the performance of the GDSC method dramatically, one approach is introducing pseudo potentials, which can avoid singularities. In our expectation, the performances of the DSC method are expected to be competitive when compared to other real-space methods such as the FD method and the FE method.